\documentclass[referee]{aa} 
\usepackage{graphicx}
\usepackage{natbib}
\usepackage{epsfig}
\bibpunct{(}{)}{;}{a}{}{,} 
\usepackage{txfonts}
\begin{document}

   \title{Near-infrared spectroscopy of 1999 JU$_{3}$, the target of the Hayabusa 2 mission (Research Note)}

   \author{N.~Pinilla--Alonso
          \inst{1}
          \and
	V.~Lorenzi\inst{2}
	\and
	H.~Campins\inst{3}
	\and
	J.~de Leon\inst{4}
	\and
	J.~Licandro\inst{5,6}
          }

   \institute{Instituto de Astrof\'isica de Andaluc\'ia- CSIC, Glorieta de la Astronom\'ia, s/n. E-18008, Granada--Spain\\
              \email{npinilla@utk.edu}
         \and Fundaci\'on Galileo Galilei -- INAF, Rambla Jos\'e Ana Fern\'andez P\'erez, 7, 37812, Bre\~na Baja, TF--Spain
         \and University of Central Florida,Physics Department, PO Box 162385, Orlando, FL, 32816--2385, USA
         \and Departamento de Edafolog\'ia y Geolog\'ia, Universidad de La Laguna, Avda. Astrof\'isico Francisco S\'anchez, s/n, 38205, La Laguna, TF--Spain
	\and Instituto de Astrof\'isica de Canarias (IAC), C/ V\'ia L\'actea s/n, 38205, La Laguna, TF--Spain
         \and Departamento de Astrof\'isica, Universidad de La Laguna, E-38205, La Laguna, Tenerife, Spain
	}

   \date{Received December 2012; accepted 4 March 2013}

 
  \abstract
   {Primitive asteroids contain complex organic material and ices relevant to the origin of life on Earth. These types of asteroids are the target of several-sample return missions to be launched in the next years. 1999 JU$_{3}$ is the target of the Japanese Aerospace Exploration Agency's Hayabusa 2 mission.}
   {1999 JU$_{3}$ has been previously identified as a C-class asteroid. Spectroscopic observations at longer wavelengths will help to constrain its composition.}
   {We obtained spectroscopy of 1999 JU$_{3}$ from 0.85 to 2.2 $\mu$m, with the 3.6 m Telescopio Nazionale Galileo using the low resolution mode of the Near Infrared Camera Spectrograph.}
   {We present a near-infrared spectrum of 1999 JU$_{3}$ from 0.85 to 2.2 $\mu$ that is consistent with previously published spectra and with its C-type classification.}
   {Our spectrum confirms the primitive nature of 1999 JU$_{3}$ and its interest as target of the sample-return mission Hayabusa 2.}


   \maketitle
%

\section{Introduction}
\begin{figure*}
   \centering
   \includegraphics{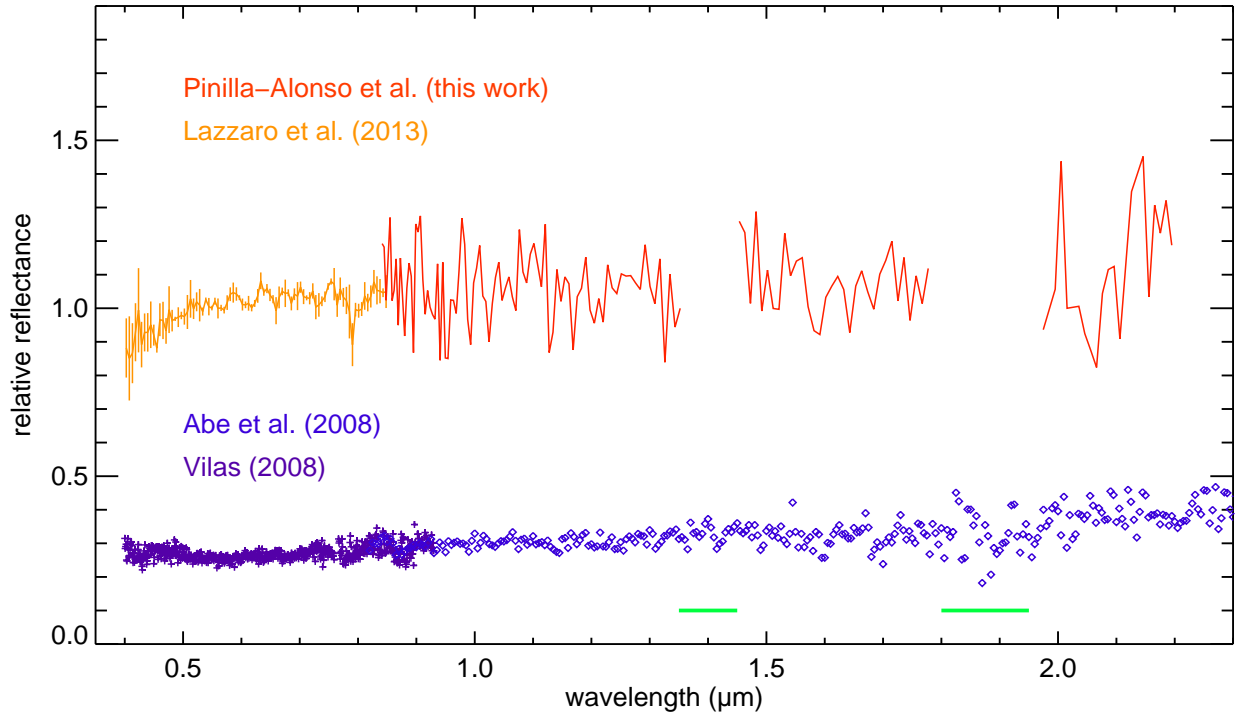}
   
   \caption{Spectrum presented in this work merged with the averaged visible spectrum from \cite{laz13} and compared with the composite spectrum from \cite{vil08} and \cite{abe08}. They are normalized at 0.55 $\mu$m and the one merged using \cite{vil08} and \cite{abe08} has been shifted for clarity. The green horizontal bars are the wavelengths dominated by telluric absorptions.}
   \label{Spectrum}
    \end{figure*}
    
Asteroids with a low albedo are believed to be composed of the most primitive materials. They are found in the B, C, D, F, G, and P classes in the Tholen classification system \citep{thobar89}. Primitive asteroids are present in the inner belt (within 2.5 AU) but are more abundant in the middle and outer main belt. They are expected to contain complex organic materials, which have been identified from spectroscopic observations on the surface of several asteroids in the outer belt \citep{cam10,riveme10,lic11}.
So far, no space mission has visited a primitive asteroid. Sample-return missions allow us to analyze the samples from solar system objects in terrestrial laboratories, thereby obtaining measurements that cannot be performed in situ. The possibility offered by sample-return missions of identifying and analyzing the structure of the organic matter in the laboratory with a high level of detail is crucial to determine whether these objects delivered the prebiotic material that led to the formation of life on Earth. Hence, for  the Japanese Aerospace Exploration Agency's Hayabusa 2, the NASA OSIRIS-REx mission and the European Space Agency MarcoPolo-R mission, samples of a primitive asteroid are highly desirable.  

Asteroid (162173) 1999 JU$_{3}$ is the target of the Hayabusa 2 mission, which has a planned launch in 2015 with the return of the asteroid samples by 2020. According to \citet{cam12}, this near-Earth Asteroid (NEA) probably escaped from the inner belt through the $\nu_{6}$ resonance from a background group of asteroids with low albedo and low inclinations.

1999 JU$_{3}$ was first classified as a Cg-type based on visible spectroscopy \citep{bin01} in the Small Main-Belt Asteroid Spectroscopic Survey classification system (SMASS; \citealt{busbin02a,busbin02b}). Cg-type asteroids typically show a pronounced UV/blue absorption starting near an upper wavelength of 0.55 $\mu$m, a relatively flat spectrum across the 0.55 to 0.9 $\mu$m wavelength range, and occasionally a small amount of absorption beginning near 0.9 $\mu$m. \citet{vil08} presents two visible spectra of 1999 JU$_{3}$ collected on different nights between July and September 2007. They have a different signal-to-noise ratio (S/N), but none of them  shows absorption in the near-ultraviolet wavelengths. On the other hand, \citet{vil08} detected a band centered at $\sim$0.7 $\mu$m that could be due to aqueous alteration. The apparent differences between the available visible spectra might be caused by compositional heterogeneity on the surface of this object. However, recent observations at different rotational phases of 1999 JU$_{3}$ \citep{laz13}, i.e., sampling different regions of the asteroid's surface, indicate an almost uniform composition with only weak variations, and rule out any 0.7 $\mu$m feature on the spectrum of 1999 JU$_{3}$. Despite the small differences between the available visible spectra, all of them agree with the suggestion 1999 JU$_{3}$ is a primitive C-type asteroid.

Up to now, only one spectrum was available in the near-infrared \citep{abe08}. This spectrum is featureless and slightly red, in agreement with what is expected for an asteroid in the C-class \citep{dem09}. In this work we present a new near-infrared (NIR) spectrum obtained in June 2012 and compare it with the other NIR spectrum available from \cite{abe08}.


\section{Observations}
    
We obtained low-resolution spectroscopy of 1999 JU$_{3}$ on 27 June 2012 with the 3.6 m Telescopio Nazionale Galileo (TNG) using the low-resolution mode of the Near Infrared Camera Spectrograph (NICS). This is a multi-mode instrument based on a HgCdTe Hawaii 1024 x 1024 array \citep{oli00}. All spectroscopic modes use the large-field camera with a plate scale of 0.25''/pixel and a field of view of 4.2' x 4.2'. A 1.5'' width slit, corresponding to a spectral resolving power of R $\sim$ 34 that is quasi-constant along the spectrum we used. The identification of the asteroid was completed by taking a series of images through the Js filter ($\lambda_{cent}$ = 1.25 $\mu$m). We identified the asteroid as a moving object at the predicted position and with the predicted proper motion.The slit was oriented along the parallactic angle and the tracking was performed at the asteroid's proper motion. The acquisition consisted of series of short-exposure images ($T_{exp}$=90 $s$) offsetting the object between positions A and B in the slit direction by 40 pixels, corresponding to 10''. This process was repeated and a number of ABBA cycles were acquired, with a total on-object exposure time of 1440 s. We used the observing and reduction procedure described by \citet{lic06}.

To correct for telluric absorption and derive the relative reflectance, the G stars Landolt (SA) 102--1081 and Landolt (SA) 107--998 \citep{lan92} were observed at different airmasses during the night, before and after observing the asteroid. The use of these Landolt stars as solar analogs was studied by \citet{lic03}. It was found that in the near-infrared and at low spectral resolution, the spectra of these stars are similar to those of solar analog stars and can be used in the reduction process in the same way. Dividing the spectra of the solar analogs by each other and then normalizing to unity around 1.6 $\mu$m, we observed that the resulting ratio was very flat and the uncertainty in the slope was smaller than 0.4\%/1000 $\AA$, meaning that the atmosphere was very stable during the observations.

After a standard flat field correction and extraction of the spectra, all AB pairs of 1999 JU$_{3}$ were averaged and divided by the spectra of the two solar analog stars. The resulting spectra where averaged into one final spectrum. In Fig. 1 the resulting near-infrared spectrum is presented and compared with the other spectrum from \citet{abe08}. Around the two strong telluric water band absorptions, the S/N of the spectrum is very low. Even in a rather stable atmosphere, the telluric absorption can vary between the object and solar analog observations and introduce artifacts. Therefore, any spectral feature detected within the 1.35--1.46 and 1.82--1.96 $\mu$m regions cannot be considered real. We do not represent those regions in our spectrum for clarity. Data beyond 2.2 $\mu$m are too noisy and are not presented here.


\section{Results}

In Fig. 1 we show our spectrum (0.85 -- 2.2 $\mu$m) compared with the NIR spectrum from \citet{abe08}. Both spectra are featureless and reddish as expected for C-type asteroids. To better compare these two spectra we computed the spectral slope from 0.85 to 2.2 $\mu$m. To do that we performed a linear fit to the data over the whole wavelength range. We computed a total of ten slope values, each one removing a randomly chosen 20\% of the data before the fit. The final spectral slope is the average of the ten computed slopes. The error comes from the maximum deviation of the slope from the average. We repeated the process with the spectrum from \citet{abe08}. The resulting spectral slopes are $S'_{this work}$ = 0.37 $\pm$ 0.28 $\%$/1000 $\AA$ and $S'_{Abe}$ =  0.89 $\pm$  0.03 \%/1000 $\AA$. Our spectrum is slightly less red than that from \citet{abe08}. However, considering both the error in the slope and the systematic errors, the two spectra are compatible with the spectra of C-type asteroids. Based on our observational experience, the systematic errors introduce an uncertainty in the slope of $\sim$  0.5 \%/1000 $\AA$, larger than the error in the calculation of the slope, which is related to the noise in the spectrum. We use the largest of the two as the error in the slope for the remainder of this paper.

To cover the visible wavelength range, we merged the spectrum from \citet{abe08} with that of \citet{vil08}. To extend our spectrum to the visible wavelengths we used the spectra from \cite{laz13}. In this work the authors present three new visible spectra of 1999 JU$_{3}$ that cover 70\% of the surface of the asteroid. The data show featureless C-type spectra without  indication of aqueous alteration. There are some discrepancies between these three spectra in the blue edge, but the authors cannot discard that they are the result of differential atmospheric refraction. However, all three spectra show no indication of hydrated material on the surface covered by the observations. We averaged the spectra and merged the result with our spectrum using the overall shape from 0.55 to 0.85 $\mu$m. For the representation we use a rebinned version of the averaged spectrum from \cite{laz13} with a binning of 20 points. Although our spectrum is noisy, the shape of the composite final spectrum from 0.4 to 2.2 $\mu$m is compatible with the overall shape of the Abe/Vila spectrum, with a higher S/N, as can be seen Fig. 1

Finally, from the shape of the final vis+NIR spectrum of 1999 JU$_{3}$ we can confirm that this is a C-type asteroid spectrally different from B-type asteroids, as is the case of 1996 RQ$_{36}$, the primary target of the Osiris-Rex Mission \citep{lau10}, and 2008 EV$_{5}$, the new primary target of the MarcoPolo-R mission. B-types are characterized by their visible spectrum, which is typically linear and featureless, with a negative spectral slope. Recently, \cite{del12} showed that the vis+NIR spectra of B-type asteroids are blue up to 0.9 $\mu$m and then either turn red or keep a blue slope. However, our spectrum is consistently red across the whole wavelength range, so its shape does not match that of the B-types, either in the visible or in the near-infrared.

\section{Conclusions}

We presented a new near-infrared spectrum of 1999 JU$_{3}$. This spectrum is featureless and slightly red with an spectral slope, S'$_{NIR}$ = 0.37 $\pm$ 0.5. We compared our spectrum with the other near-infrared spectrum available in the literature \citep{abe08}. Both are compatible within the S/N and are consistent with a classification of the asteroid as C-type suggesting that there is no evident heterogeneity in the surface composition of this asteroid. Consistently, all spectra available at this moment in the visible and the near-infrared result in a classification of this object as a primitive asteroid which makes it a very interesting target for the Hayabusa 2 mission. This is reinforced in particular because the tempting feature detected by \citet{vil08}, which could have been related with some degree of thermal processing, has been ruled out by new observations  that cover different areas of the surface of 1999 JU$_{3}$ \citep{laz13}.

\begin{acknowledgements}
Based on observations made with the Italian Telescopio Nazionale Galileo (TNG) operated on the island of La Palma by the Fundaci\'on Galileo Galilei of the INAF (Istituto Nazionale di Astrofisica) at the Spanish Observatorio del Roque de los Muchachos of the Instituto de Astrofisica de Canarias. NPA acknowledges  support by contract AYA2011-30106-C02-01. JdL acknowledges financial support from the current Spanish ``Secretar\'{\i}a de Estado de Investigaci\'on, Desarrollo e Innovaci\'on'' (Juan de la Cierva contract). JL acknowledges support from the project AYA2011-29489-C03-02 (MINECO).
\end{acknowledgements}

\bibliographystyle{aa}
\bibliography{bibJU3}

\end{document}